\documentclass[hyper]{JHEP} 

\usepackage{epsfig}  
\title{Exact Solutions in SFT and Marginal Deformation
in BCFT}
\author{ J. Kluso\v{n}
\footnote{On leave from Masaryk University, Brno}\\
Institute of Theoretical Physics, University of Stockholm, SCFAB\\
SE- 106 91 Stockholm, Sweden \\
and \\
Institutionen f\"or teoretisk fysik\\
BOX 803, SE- 751 08 
Uppsala, Sweden \\
E-mail: \email{josef.kluson@teorfys.uu.se}}   \preprint{\hepth{0303199}}  			  	
 \abstract{In this  note we will study 
 solution of open bosonic string field theory
based on  action of operators from chiral algebra
of boundary conformal field theory on identity element
of string field theory star algebra. We will demonstrate that the
string field theory action for fluctuation fields  around
this classical solution  can be mapped to  the string field theory action
defined through the  new   boundary conformal
field theory that arises from the original one
through the marginal deformation inserted on  the
world-sheet  boundary.}


\def\bra #1{\left<#1\right|}
\def\ket #1{\left|#1\right>}

\def\fI{f_{\mathcal{I}}}
\def\Id{\mathcal{I}}
\def\mw{\mathcal{W}}
\def\mbw{\mathbf{W}}
\def\oz{\overline{z}}
\def\oW{\overline{W}}
\begin{document}
\section{Introduction}\label{first}
Much work has focused on the search 
for classical solutions of cubic bosonic open
string field theory (SFT) \cite{Witten:1985cc}
(For review, see \cite{Ohmori:2001am,Arefeva:2001ps,
DeSmet:2001af,Taylor:2002uv}.). Despite important technical
progress in the understanding of the open string
star product-notably the discovery of a
new connections with non-commutative field theory
\cite{Bars:2003cr,Bars:2001ag,Rastelli:2001hh,
Bars:2002bt,Bars:2002nu,Douglas:2002jm,
Belov:2002fp,Arefeva:2002jj,Bars:2002yj,Bars:2002qt}
it is still very difficult to find analytic classical solution.
Exception is the special form of 
SFT, vacuum string field theory  (VSFT), where
 exact results have been obtained
\cite{Rastelli:2001jb,Kostelecky:2000hz,
Rastelli:2001rj,Rastelli:2001vb,Gross:2001rk,
Gross:2001yk,Schnabl:2002gg,Bonora:2002pu,
Bonora:2002iq}.

 In our recent  papers \cite{Kluson:2002av,Kluson:2002kk,Kluson:2002ex,
Kluson:2002hr,Kluson:2002gu}
we found  some  exact solutions of 
SFT equation of motion
\footnote{For closely
related papers, see 
\cite{Takahashi:2001pp,Takahashi:2002ez,Kishimoto:2002xi,Takahashi:2003pp}.
For other papers considering exact solutions in SFT, see 
 \cite{Kling:2002vi,Kling:2002ht,Lechtenfeld:2002cu,Lechtenfeld:2000qj}.}.
All these solutions have the property  that when
we expand the string field
 around this solution and insert it into the  
SFT action then the  
 new BRST operator $Q'$ 
suggests that SFT action  for fluctuation modes
is in some sense related   to  the SFT action defined around the
new  background boundary conformal field theory ($BCFT''$)
 that arises from
the original  $BCFT$  by presence
of marginal deformation on the world-sheet boundary.
The possible relation between these two actions 
will be indication of the background independence 
of SFT 
\cite{Sen:1990hh,Sen:1990na,Sen:1992pw,
Sen:1993mh,Sen:1994kx}  which unfortunatelly is not
explicitly seen in its formulation.  Let us say more about this issue.

This issue arises because SFT can be written after choosing what
amounts to a classical solution, namely $BCFT$ with central charge
$c=0$. As a result the SFT actions $S_1\ , S_2 $ written
using two different $BCFT's$ $BCFT_1$ and 
$BCFT_2$ are not manifestly equivalent. However it was shown in
\cite{Sen:1990hh,Sen:1990na,Sen:1992pw,
Sen:1993mh,Sen:1994kx} that for the case when 
  $BCFT_1$ and 
$BCFT_2$ are nearby theories related by infinitesimal 
marginal deformation, one can prove that $S_1$ and
$S_2$ are the same action expanded around different 
solutions. 

In this paper we will try to study this problem
from slightly different point of view. In particular,   we focus on 
the  relation between  solution of SFT equation of
motion given in \cite{Kluson:2002av,Kluson:2002kk,Kluson:2002ex,
Kluson:2002hr,Kluson:2002gu}
 and the marginal  deformation in  $BCFT$. 
Our goal is to show that when we   expand  string
field around classical solution and insert it  into
 the original
SFT action $S$ which is defined on background $BCFT$, we obtain after suitable
redefinition of the fluctuation modes  
the SFT action $S'$  defined on $BCFT''$  that
is related to the original $BCFT$  by inserting marginal
deformation on the boundary of the world-sheet 
(For more details about marginal deformations in
$BCFT$, see
\cite{Recknagel:1998ih}.). To say differently,
we will show that two SFT actions $S, \  S'$ written
using two different $BCFT \ , BCFT''$  which are 
 related by marginal deformation, are in fact two SFT
actions expanded around different classical solutions. 

In order to show this equivalence  will consider 
 operators, that determine exact solution of
SFT, which  
belong to the chiral algebra $\mathcal{W}$ of 
$BCFT$
\footnote{For very nice review to the subject of 
$BCFT$ see for example
\cite{Gaberdiel:2002my,Schomerus:2002dc} and reference therein.}.
More precisely,  we will construct the solution of SFT equation of motion
that is based on an action of some operator $\mbw$
from $\mathcal{W}$  on the SFT star algebra
 identity element $\mathcal{I}$.
Then we will study the fluctuation modes around this
solution. It turns out that after performing suitable redefinition
of the fluctuation modes  we will  get  correlation 
functions  
\footnote{These $BCFT$ correlation functions  
 define SFT action $S$  in the CFT description
\cite{LeClair:1988sp,LeClair:1988sj}.}
which have the same form as the correlation functions in the deformed
$BCFT''$ that arises from the original one by insertion of
marginal interaction $\mbw(x)$ on the  world-sheet boundary.
In order to be able to perform such a identification
we took  the operators $\mbw$ from $\mw$ 
since only in that case we have
 well defined  deformed correlation functions
in $BCFT$ containing bulk and boundary operators
as well as operator $\mbw(x)$ 
\cite{Recknagel:1998ih}. 
  In other words we will show
that the SFT action $S'$  for fluctuation modes is the same
as the SFT action defined by the second $BCFT''$. This result can be considered
as more precise extension  of  the analysis given in 
\cite{Kluson:2002av} in the sense that we explicitly
show the equivalence of these two actions. 
We mean that this result could be considered as 
further indication of  the  background independence
in SFT \cite{Sen:1990hh,Sen:1990na,Sen:1992pw,
Sen:1993mh,Sen:1994kx}.

The organization of this paper  is as follows. In the
next section (\ref{second}) we review the basic facts about
SFT, bulk an boundary CFT. Then we turn to the construction
of the classical solution of SFT equation of motion following
 \cite{Kluson:2002av,Kluson:2002kk,Kluson:2002ex,
Kluson:2002hr,Kluson:2002gu}.
In section (\ref{third}) we will study  
fluctuation modes above the classical solution
and we will find the precise relation between
the SFT $S'$  for fluctuation modes and SFT defined 
on $BCFT''$. 
In conclusion (\ref{fourth}) we will  outline our results. 
\section{General solutions}\label{second}
We begin this section with the review  the basic facts about
bosonic string field theory, following
mainly \cite{Ohmori:2001am,
Arefeva:2001ps,DeSmet:2001af}. Gauge invariant
string field theory is described with the
full Hilbert space of the first quantized open
string including $b,c $ ghost fields subject
to the condition that the states must
carry ghost number one, where $b$ has ghost
number $-1$, $c$ has ghost number $1$ and
$SL(2,C)$ invariant vacuum $\ket{0}$
carries ghost number $0$. We denote
$\mathcal{H}$ the subspace of the full Hilbert
space carrying ghost number $1$.  Any state
in $\mathcal{H}$ will be denoted as $\ket{\Phi}$
and corresponding vertex operator $\Phi$
is the  vertex operator that  creates state $\ket{\Phi}$
out of the vacuum state $\ket{0}$
\begin{equation}
\ket{\Phi}=\Phi \ket{0} \ .
\end{equation}
Since we are dealing with open string theory,
the vertex operators should be put on the boundary
of the world-sheet
\footnote{Since these states describe open string they 
belong to the class so-named boundary operators
in $BCFT$ that live on
the real line $y=0$ for $z=x+iy$ \cite{Schomerus:2002dc}.}. 
  The string field theory action
is given
\cite{Witten:1985cc}
\begin{equation}\label{actionCFT}
S=-\frac{1}{g_0^2}\left(
\frac{1}{2\alpha'}\left<I\circ \Phi (0) Q \Phi (0)\right>+
\frac{1}{3}\left<f_1\circ \Phi(0) f_2\circ \Phi(0)
f_3\circ \Phi(0)\right>\right) \ ,
\end{equation}
where $g_0$ is open string coupling constant, $Q$
is BRST operator and $<>$ denotes correlation function
in the combined matter ghost conformal field theory in
the upper half plane $ \mathrm{Im} z\geq 0$.
$I,f_1, f_2, f_3$ are conformal mapping  exact
form of which is reviewed in \cite{Ohmori:2001am} and $f_i\circ\Phi(0)$ denotes
the conformal transformation of $\Phi(0)$ by $f_i$. For
example, for $\Phi$ a primary field of dimension $h$,
then $f_i\circ \Phi(0)=(f'_i(0))^h\Phi(f_i(0))$. 
In  our calculation we use convention from
the very nice review
\cite{Ohmori:2001am}
\begin{eqnarray}\label{conM}
T_m(z)=-\frac{1}{\alpha'}\partial_z X^{\mu}_L(z)
\partial_z X^{\nu}_L\eta_{\mu\nu} 
 \ , \nonumber \\
\overline{T}_m(\overline{z})=
-\frac{1}{\alpha'}\partial_{\overline{z}}
 X^{\mu}_R(\overline{z})
\partial_{\overline{z}} 
X^{\nu}_R(\overline{z})\eta_{\mu\nu} 
 \ , \nonumber \\
 X^{\mu}_L(z)X^{\nu}_L(w)\sim
-\frac{\alpha'}{2}\eta^{\mu\nu}
\ln (z-w)  \ , \nonumber \\
 X^{\mu}_R(\overline{z})
X^{\nu}_R(\overline{w})\sim
-\frac{\alpha'}{2}\eta^{\mu\nu}
\ln (\overline{z}-\overline{w})  \ , 
\nonumber \\
X^{\mu}_L(z)X^{\nu}_R(\overline{w})\sim
-\frac{\alpha'}{2}\eta^{\mu\nu}
\ln (z-\overline{w})  \  
\nonumber \\
\end{eqnarray}
with the BRST operator 
\begin{eqnarray}
Q=\frac{1}{2\pi i} \int_C
dz j_B(z)-\frac{1}{2\pi i}\int_C d
\overline{z}\overline{j}_B(\overline{z}) \ , \nonumber \\
j_B(z)=c(z)
\left[T_m(z)+\frac{1}{2}T_{gh}(z)\right] \ ,
\nonumber \\
\overline{j}_B(\overline{z})=
\overline{c}(\overline{z})
\left[\overline{T}_m(
\oz)+
\frac{1}{2}\overline{T}_{gh}(
\oz)\right] \ ,
\nonumber \\
\end{eqnarray}
where $j_B(z)$ is holomorphic and $\overline{j}_B(\oz)$
 is anti-holomorphic
current  and where $T_{ghost}\ ,\overline{T}_{ghost}$ are
holomorphic and anti-holomorphic  stress energy tensor
for the ghosts. 
In what follows we will not need
to know the  explicit form of the ghost contribution.

It turns out   that the natural definition of
the string field theory is in language of 
$BCFT$ \cite{LeClair:1988sp,LeClair:1988sj}.
$BCFT$  are usually regarded  as associated boundary theories
to the bulk CFT theories. Recall that bulk CFT are
defined on the whole complex plane and they appear in
the world-sheet description of the closed strings. Their
state spaces $\mathcal{H}^P$ contain all the closed
string modes and the coefficients $C=C^P$ of their operator
product expansions  (OPE) encode closed string interactions.
The space $\mathcal{H}^P$ is equipped with the
action of a Hamiltonian $H^P$ and of field
operators $\phi(z,\overline{z})$. According to
state-field correspondence we have an identification
\begin{equation}
\phi(z,\overline{z})=\Phi^P(\ket{\phi},z,\overline{z}) \ ,
\mathrm{for \ all } \ket{\phi} \in \mathcal{H}^P \ .
\end{equation}
Among the fields of a CFT one distinguishes so-called
chiral fields which depend on only one of the coordinates
$z$ or $\overline{z}$ so that they are either holomorphic,
$W=W(z)$, or anti-holomorphic, $\overline{W}=\overline{W}(
\overline{z})$. The (anti)-holomorphic fields of given theory,
or their Laurent modes $W_n \ , \overline{W}_n$ defined
through 
\begin{equation}
W(z)=\sum W_nz^{-n-h} \ , 
\oW(\oz)=\sum \oW_n\oz^{-n-\overline{h}} \ ,
\end{equation}
generate two commuting chiral algebras $\mathcal{W}$ and
$\overline{\mathcal{W}}$. 
 The most important of these
chiral fields, the Virasoro fields $T(z), \overline{T}(\overline{z})$
with modes $L_n \ , \overline{L}_n$ are among
the chiral fields and  numbers $h\ , \overline{h}$ are the (half-)integer
conformal weights of $W(z)\ , \oW(\oz)$.

 $BCFT$  are conformal field theories on the upper
half-plane $\mathrm{Im} z \geq 0$ which in the interior $\mathrm{Im} >0$ 
are locally equivalent to the given bulk theory: The
state space $\mathcal{H}^H$ of the $BCFT$ is equipped with the action
of a Hamiltonian $H^H$ and of bulk fields 
\begin{equation}
\phi(z,\overline{z})=\Phi^H(\ket{\phi},z,\overline{z}) \ ,
\end{equation}
which are assigned to the same fields as were used to
label filed in the bulk theory. However these
fields $\phi$ act on a different space of states
$\mathcal{H}^H$. We also demand that all the leading
terms in the OPE's of bulk fields coincide with
the OPE's in the bulk theory. Having the same singularities
as in the bulk theory means that the boundary conditions do not affect
the equation of motion in the bulk. 
We must also require the boundary theory to be conformal.
This is guaranteed if the Virasoro field obeys
the following gluing condition
\begin{equation}
T(z)=\overline{T}(\overline{z}) \ ,
z=\overline{z} \ .
\end{equation}
We also presume that all chiral fields $W(z),
\overline{W}(\overline{z})$ can
be extended analytically to the real
line and that there exist a local automorphism
$\Omega$-called the gluing  map of the chiral
algebra  $\mathcal{W}$ such that
\begin{equation}\label{gluing}
W(z)=\Omega \left(\overline{W}\right)(\overline{z})  
\ , z=\overline{z} \ .
\end{equation}
Now the assumption of the existence of
the gluing map $\Omega$ has powerful 
consequence that it gives rise to an
action of one chiral algebra $\mathcal{W}$ 
on the state space $\mathcal{H}^H$ of the boundary
theory. More precisely, we combine
the chiral fields $W(z)$ and
$\Omega \overline{W}(\overline{z})$ into single
object $\mathbf{W}(z)$  defined on the
whole complex plane such that
\begin{equation}
\mathbf{W}(z):= W(z) \ , \mathrm{Im} z\geq 0 \ ,
\mathbf{W}(z):=\Omega \overline{W}(\overline{z})
\ , \mathrm{Im} z<0 \ .
\end{equation}
Thanks to the gluing condition along the boundary 
this field is analytic and we can expand it in
a Laurent series 
\begin{equation}
\mathbf{W}(z)=
\sum_n W_nz^{-n-h}
\end{equation}
so that we introduce the modes $W_n$ that 
acts on the Hilbert space $\mathcal{H}^H$. Then we can
obtain the modes $W_n$ through the
integration over the curve in the upper half
complex plane
\begin{equation}
W_n=\frac{1}{2}\left(\frac{1}{2\pi i}\int_C dz z^{n+h-1}
W(z)-\frac{1}{2\pi i}\int_C d\overline{z}
\overline{z}^{n+h-1}\Omega (\overline{W})(
\overline{z})\right) \ ,
\end{equation}
where $C$ is curve in the upper half plane that
is labeled as $z=-e^{-i\sigma +\tau} \ ,
\sigma \in (0, \pi)$. 
It is important to stress that there is just one
such action of $\mathcal{W}$ constructed out
of the two chiral fields $W(z)$ and $\Omega \overline{W}(\overline{z})$.

After this short review of $BCFT$ we return
to the SFT  and its
equation of motion. Note that in  the abstract language 
\cite{Witten:1985cc}
the  open string field theory action (\ref{actionCFT})
is
\begin{equation}\label{actionW}
S=-\frac{1}{g_0^2}\left(
\frac{1}{2\alpha'}\int \Phi \star Q\Phi
+\frac{1}{3}\int \Phi\star\Phi\star\Phi\right) \
\end{equation}
from which we immediately get an equation of motion
\begin{equation}\label{eq}
\frac{1}{\alpha'}Q\Phi_0+\Phi_0\star \Phi_0=0  \ .
\end{equation}
It is easy to see that the string field in the form
\begin{equation}\label{Proll}
\Phi_0=e^{-\lambda K_L(\mathcal{I})}\star \frac{1}
{\alpha'}Q( e^{\lambda K_L(\mathcal{I})}) \ , 
\lambda \in \mathrm{R}
\end{equation}
is solution of (\ref{eq}) for any ghost number zero
operator $K_L$ acting on the SFT star  algebra $\star$
 identity element  $\mathcal{I}$ which is ghost number zero
string field that obeys
\cite{Horowitz:dt} 
\begin{equation}
\mathcal{I}\star X=X\star \mathcal{I}=X \ ,
\end{equation}
for any string field $X$
\footnote{For recent study of the identity element
$\mathcal{I}$, see
\cite{Ellwood:2001ig,Matsuo:2001yb,Kishimoto:2001ac,
Schnabl:2002gg,Kishimoto:2001de}.}.
Let us consider   operator $K$
 in (\ref{Proll})
from  the chiral algebra $\mw$ of
$BCFT$ in the form 
\begin{equation}\label{K}
K\equiv \mbw=\frac{1}{2}\left(\frac{1}{2\pi i}\int_C
dz W(z)-\frac{1}{2\pi i}\int_C d\oz \Omega 
(\overline{W})(\oz)\right) \ . 
\end{equation}
where  $W(z), \overline{W}(\oz)$ are holomorphic,
anti-holomorphic fields respectively of
conformal weight  $(1,0)\ , (0,1)$  which
 transform under general conformal transformations
$z\rightarrow f(z)$ as
\begin{eqnarray}
 U_f W(z)U_f^{-1}=\frac{df(z)}{dz}W(f(z))\equiv
f'(z)W(f(z)) \ \ , \nonumber \\
U_f \overline{W}(\oz)U_f^{-1}=\frac{d\overline{f}
(\oz)}{d\oz}\overline{W}(\overline{f}(\oz))\equiv
\overline{f}'(\oz)\oW(\overline{f}(\oz))  \ . \nonumber \\
\end{eqnarray}
Then we immediately get that $\mbw$ is invariant under
conformal transformations
\begin{eqnarray}\label{conW}
f\circ \mbw\equiv
U_f\mbw U_f^{-1}=
\frac{1}{2\pi i}\int_C dz f'(z)
W(f(z))-\frac{1}{2\pi i}
\int_C d\oz \overline{f}'(\oz)
\oW(\overline{f}(\oz))
=\nonumber \\
=\frac{1}{2\pi i}\int_{f(C)} dz f'(z)
W(f(z))-\frac{1}{2\pi i}
\int_{f(C)} d\oz \overline{f}'(\oz)
\oW(\overline{f}(\oz))
=\mbw \nonumber \\
\end{eqnarray}
using the fact that $\mbw$ does not depend on
the integration contour $C$ as a  consequence of
gluing conditions (\ref{gluing}).

We can also define an action 
of $\mbw$ on identity field directly in CFT language, following
 \cite{Kishimoto:2001de,Ohmori:2002kj,Ohmori:2002ah}.
Instead to taking $\ket{\mathcal{I}}$ as identity
element of star algebra we will define its through the
relation
\begin{equation}
\bra{\mathcal{I}}\left.\mathcal{O}\right>=
\left<\fI\circ \mathcal{O}\right> \ , 
\ket{\mathcal{O}}=\mathcal{O}(0)\ket{0} \ , 
\end{equation}
where 
\begin{equation}
\fI=h^{-1}(h(z)^2) \ , 
h=\frac{1+iz}{1-iz} \ .
\end{equation}
In this approach 
the identity field is considered as state that belong to family
of wedge states
\cite{LeClair:1988sp,Rastelli:2000iu,Rastelli:2001vb}.
 Wedge state $\ket{n}$  of an angle  
$\frac{2\pi}{n}$ is defined 
\begin{equation}
\left<n\right.\ket{\mathcal{O}}=
\left<f_n \circ \mathcal{O}\right> \ ,
f_n=h^{-1}\left(h(z)^{2/n}\right) \ .
\end{equation}
It follows that $\ket{\mathcal{I}}$ is 
the wedge state $\ket{n=1}$ of an angle
$2\pi$.
In this description we define action of the operator
$\mbw$ on $\Id$ as 
\begin{equation}
\bra{\Id}\mbw \ket{\phi}=\left<\fI  \circ
\mbw \fI \circ \phi(0)\right> \ .
\end{equation}
Using invariance of  $\mbw$ 
under conformal transformation we immediately get 
that $\mbw$ annihilates any wedge state since
\begin{eqnarray}
\bra{n}\mbw\ket{\phi}=
\left<f_n\circ\left(\mbw \phi(0)\right)\right>=
\left<\mbw f_n\circ\phi(0)\right>=\nonumber \\
=\left<
\frac{1}{2\pi i} \oint_{C}
dz \mbw(z)
f_n\circ \phi(0)\right>=0 \  \nonumber \\
\end{eqnarray}
by deforming the contour $C$ until it shrinks
to a point at infinity where there is no
other operator. In the upper expression 
we  used 
standard doubling trick
to express  $\mbw$ through
holomorphic current $\mbw(z)$. 
Let us apply this result for identity field
$\mathcal{I}$ and write $\mbw=\mbw_L+\mbw_R$, where
subscripts $L,R$ denote the integrals of
holomorphic and anti-holomorphic currents
$W(z)\ , \oW(\oz)$ over left and right side of
the string respectively. Then we immediately get
\begin{equation}
\mbw(\mathcal{I})=0 \Rightarrow 
\mbw_L(\mathcal{I})=- \mbw_R(\mathcal{I}) \ . 
\end{equation}
For our next purposes it is  important following
``partial integration formula'' 
\begin{equation}\label{derK}
\mbw_L (A)\star B+
 A\star \mbw_R (B)=0 \ ,
\end{equation}
where  $A \ , B$ are general string fields. 
Recent very nice discussion of
the upper expression can be found in \cite{Kishimoto:2001de},
where instead of operator $\mbw$ the BRST operator
$Q$ is considered. 
The proof given there can be easily applied for general
operator that is invariant under conformal 
transformations so that (\ref{derK}) is valid
for $\mbw$ too.

Let us return to the  solution of SFT
(\ref{Proll}). We observe  that it has the form of 
pure gauge. This fact certainly deserves
deeper explanation.  As is well known the
  string field theory action
(\ref{actionW})  is invariant 
under  small gauge transformations 
\begin{equation}
\delta \Phi=Q\Lambda-\Lambda \star \Phi+
\Phi\star \Lambda \ ,
\end{equation}
where $\Lambda $ is ghost number zero string field.
 On the other hand the action (\ref{actionW})   is not generally
invariant under the large gauge transformations 
\begin{equation}
\Phi'=e^{-\Lambda}\star Q(e^{\Lambda})+
e^{-\Lambda}
\star \Phi\star e^{\Lambda} \ .
\end{equation}
As is well known there is a sharp distinction between
the small gauge transformations and the large ones, 
for very nice discussion, 
see  \cite{Harvey:1996ur}. As was argued there,
 small gauge transformation describes redundancy in
our description of the theory. 
On the other hand, large
gauge transformations are true symmetries that relate
different solutions in given gauge theory which in our
case is the open bosonic  string field theory. We will
see that this interpretation of the large
gauge transformation is the appropriate one in
case of (\ref{Proll}). To support this claim
let us start to study  fluctuation
modes around $\Phi_0$.  
As usually we expand  string field $\Phi$  as
\begin{equation}
\Phi=\Phi_0+\Psi \ 
\end{equation}
and insert it  in (\ref{actionW}).
Then we obtain an
action for the fluctuation field $\Psi$ in
the same form as  the original one
(\ref{actionW}) 
\begin{equation}\label{actionWF}
S'=-\frac{1}{g_0^2}\left(
\frac{1}{2\alpha'}\int \Psi \star Q'\Psi
+\frac{1}{3}\int \Psi\star\Psi\star\Psi\right) \ , 
\end{equation}
where    the new  BRST operator
$Q'$ was introduced 
\cite{Horowitz:dt}
\begin{equation}\label{Q'}
Q'(X)=Q(X)+\alpha'\Phi_0\star X-\alpha'
(-1)^{|X|}X\star \Phi_0 \ .
\end{equation}
In order to obtain the new form of the BRST
operator (\ref{Q'}) we will follow the calculation
given  in \cite{Kluson:2002kk}. 
We start with the function
\begin{equation}
F(t)=\frac{1}{\alpha'}
e^{-\lambda\mbw_L(\mathcal{I})t}\star
Q(e^{\lambda\mbw_L(\mathcal{I})t}) \ \ ,
F(1)=\Phi_0 \ , F(0)=0 
\end{equation}
and perform Taylor expansion around the
point $t=1$
\begin{equation}
\Phi_0=F(1)=F(0)+
\sum_{n=1}^{\infty}
\frac{1}{n!}
\frac{d^nF}{d^nt}(0) \ ,
\end{equation}
where
\begin{eqnarray}
\frac{dF}{dt}=\frac{\lambda}{\alpha'}
e^{-\mbw_L(\mathcal{I})t}
\star [Q,\mbw]_L(\mathcal{I})\star 
e^{\mbw_L(\mathcal{I})t} \ , 
\nonumber \\
\frac{dF}{dt}(0)=\frac{\lambda}{\alpha'}
[Q,\mbw]_L(\mathcal{I})\equiv \lambda
D_L(\mathcal{I}) \ , 
\nonumber \\
\frac{d^2F}{d^2t}=\lambda
e^{-\mbw_L(\mathcal{I})t}(-\mbw_L(\mathcal{I})
\star D_L(\mathcal{I})+
D_L(\mathcal{I})\star \mbw_L(\mathcal{I})
)\star e^{\mbw_L(\mathcal{I})t} 
\ , \nonumber \\
\frac{d^2F}{d^2t}(0)=-\lambda\mbw_L(\mathcal{I})
\star D_L(\mathcal{I})+\lambda
D_L(\mathcal{I})\star\mbw_L(\mathcal{I})=
\lambda[\mbw,D]_L(\mathcal{I}) \ , \nonumber \\ 
\frac{d^3F}{d^3t}(0)=
\lambda^2[\mbw_L,[\mbw_L,D_L]]  
\ , 
\dots  \ ,
\frac{d^nF}{d^nt}(0)=\lambda^n\overbrace{[\mbw,[\mbw,\dots,
[Q,\mbw]]]_L}^{n-1}
(\mathcal{I}) 
\nonumber \\
\end{eqnarray} 
and consequently
\begin{equation}\label{phi}
\Phi_0=\frac{1}{\alpha'}
\sum_{n=1}^{\infty}
\frac{\lambda^n}{n!}\overbrace{[\mbw,[\mbw,\dots,[Q,\mbw]]]_L}^{n}
(\mathcal{I})\equiv 
\mathcal{D}_L(\mathcal{I})\ .
\end{equation}
It is important to stress that for validity of
the calculation given above $\mbw$ must obey 
the relation (\ref{derK}). We have
also  used $Q(\mathcal{I})=0 \Rightarrow
Q_R(\mathcal{I})=-Q_L(\mathcal{I}) \ , [Q_R,\mbw_L]=0$.  
From  (\ref{phi}) 
 see that we can express $\Phi_0$ 
as a result of the action 
of the ghost number one operator
$\mathcal{D}_L$ acting on the identity field. 
Then we immediately
obtain
\begin{eqnarray}\label{Qgen}
Q'(X)=Q(X)+\alpha'\mathcal{D}_L(\mathcal{I})
\star X-\alpha'(-1)^{|X|}X\star \mathcal{D}_L
(\mathcal{I})= \nonumber \\
=Q(X)-\alpha'\mathcal{I}\star 
\mathcal{D}_R(X)-\alpha'\mathcal{D}_L(X)
\star \mathcal{I}=
Q(X)-\alpha'\mathcal{D}(X)=\nonumber \\
=Q(X)+\sum_{n=1}^{\infty}
\frac{\lambda^n}{n!}\overbrace{[\mbw,[\mbw,\dots,[\mbw,Q]]]_L}^{n}
(X)=e^{\lambda\mbw}(Q(e^{-\lambda\mbw}(X))) \ . 
\nonumber \\
\end{eqnarray}
This form of the shifted BRST operator $Q'$ is
 convenient for the analysis of 
fluctuation modes around solution
(\ref{Proll}) as we show in the next
section. 
\section{Relation between  SFT action
$S'$ and the deformation in $BCFT$ }\label{third}
In the previous section we have found an exact
solution of the string field theory and
also 
an action for fluctuation modes 
\begin{eqnarray}\label{actionWF1}
S'=-\frac{1}{g_0^2}\left(
\frac{1}{2\alpha'}\int \Psi \star Q'\Psi
+\frac{1}{3}\int \Psi\star\Psi\star\Psi\right)=
\nonumber \\
=-\frac{1}{g_0^2}\left(
\frac{1}{2\alpha'}\left<I\circ \Psi (0) Q' \Psi (0)\right>+
\frac{1}{3}\left<f_1\circ \Psi(0) f_2\circ \Psi(0)
f_3\circ \Psi(0)\right>\right) \ ,
\nonumber \\
Q'(\Psi(0))=e^{\lambda\mbw}Qe^{-\lambda\mbw}(\Psi(0)) \ .
\nonumber \\ 
\end{eqnarray}
Upper expression  implies that it
is natural to  consider  following redefinition
of fluctuation states 
\begin{equation}\label{flu}
\ket{\Psi}=e^{\lambda\mbw}\ket{\Phi} \ , 
\ket{\Psi}=\Psi(x=0)\ket{0} \ ,
\ket{\Phi}=\Phi(x=0)\ket{0} \ ,
\end{equation}
where $\Psi(x) \ ,\Phi(x)$ are boundary operators
in $BCFT$ which are localized at point $x$ on 
the real line. 
Using the fact that $\mbw$ annihilates vacuum
state $\ket{0}$ we can write
\begin{equation}
\mbw\ket{\Psi}=\mbw\Psi(0)\ket{0}=
[\mbw,\Psi(0)]\ket{0} 
\end{equation}
hence 
\begin{equation}
\ket{\Psi}=e^{\lambda\mbw}\ket{\Phi}=
\sum_{N=0}^{\infty}\frac{\lambda^N}
{N!}\overbrace{
[\mbw,[\mbw,\dots,[\mbw,\Phi(0)]]]}^{N}\ket{0} \ ,
[\mbw,\Phi(0)]=\frac{1}{2\pi i}\oint_C dz \mbw(z)
\Phi(0)  
\end{equation}
so that we can define 
the vertex operator for fluctuation field as
\begin{equation}\label{redef}
\Psi(x)=e^{\mbw}(\Phi)(x)\equiv
\sum_{N=1}^{\infty}
\frac{\lambda^N}{2^NN!}\oint_{C_1}\dots
\oint_{C_N}
\frac{dz_1}{2\pi i}\dots \frac{dz_N}
{2\pi i} 
W(z_1)\dots W(z_N)\Phi(x) \ ,
\end{equation}
where $C_i$ are small circles around the point
$x$ and where their radii are given as
$\epsilon_{i-1}>\epsilon_{i}$ and in the end
of the calculation we take the limit $\epsilon_i
\rightarrow 0$.
In the previous expression we have slightly moved
the insertion point $x$ above to real axis in order
to perform contour integration. As a result the
second term in $\mbw$ has no singularity with $\Phi(x)$ and
hence we can consider the holomorphic field
$W(z)$ only. 
   
When we insert (\ref{flu}) into 
(\ref{actionWF1}) we get 
\begin{eqnarray}\label{actionWF2}
S=-\frac{1}{g_0^2}\left(
\frac{1}{2\alpha'}\int e^{\lambda\mbw}(\Phi) \star 
e^{\lambda\mbw}Q(e^{-\lambda\mbw}
e^{\lambda\mbw}(\Phi))
+\frac{1}{3}\int e^{\lambda\mbw}(\Phi)\star e^{\lambda\mbw}(\Phi)
\star e^{\lambda\mbw}(\Phi)\right)=
\nonumber \\
=-\frac{1}{g_0^2}\left(
\frac{1}{2\alpha'}\left<I\circ \left(e^{\lambda\mbw}(\Phi) (0)\right)
e^{\lambda\mbw}(Q \Phi (0))\right>+\right. \nonumber \\
\left. +
\frac{1}{3}\left<f_1\circ \left(e^{\lambda\mbw}(\Phi)(0)
\right) f_2\circ \left(e^{\lambda\mbw}(\Phi)(0)\right)
f_3\circ \left(e^{\lambda\mbw}(\Phi)(0)\right)\right>\right) \ .
\nonumber \\
\end{eqnarray}
Using (\ref{conW}) we obtain
\begin{equation}
f_i \circ (e^{\lambda\mbw}(\Phi)(0))=
U_{f_i}e^{\lambda\mbw}U_{f_i}^{-1}\left(
U_{f_i}\Phi(0)U_{f_i}^{-1}\right)=e^{\lambda\mbw}(
f_i \circ \Phi(0)) \ 
\end{equation}
so that the SFT
action $S'$ for fluctuation modes can be written as
\begin{eqnarray}\label{actionWF2N}
S'=-\frac{1}{g_0^2}\left(
\frac{1}{2\alpha'}\left<e^{\lambda\mbw}
\left(I\circ \Phi(0)\right)
e^{\lambda\mbw}(Q \Phi (0))\right>+\right. \nonumber \\
\left. +
\frac{1}{3}\left< e^{\lambda\mbw}(f_1\circ\Phi(0))
  e^{\lambda\mbw}(f_2\circ\Phi(0))
 e^{\lambda\mbw}(f_3\circ\Phi(0))\right>\right)= 
\nonumber \\
=\frac{1}{g_0^2}\left(
\frac{1}{2\alpha'}\left<I\circ \Phi(0)
Q \Phi (0)\right>_{W,\lambda}+
\frac{1}{3}\left<f_1\circ\Phi(0)
  f_2\circ\Phi(0)
 f_3\circ\Phi(0)\right>_{W,\lambda}\right) \ ,
\nonumber \\
\end{eqnarray}
where we have defined deformation 
of boundary correlators \cite{Recknagel:1998ih} 
\begin{equation}\label{defm}
\left<\Phi_1(x_1)\dots
\Phi_M(x_M)\right>_{W,\lambda}=
\left<e^{\lambda\mbw}(\Phi_1)(x_1)\dots
e^{\lambda\mbw}(\Phi_M)(x_M)\right> \ .
\end{equation}
The form of the action 
(\ref{actionWF2N}) is the main result of our paper
which  says   that when we perform 
redefinition of fluctuation modes as in (\ref{redef}),
then the  SFT  $S'$
(\ref{actionWF1}) is the same as the
SFT action  defined on
the background $BCFT''$ that arises from the
original one through  marginal deformation inserted
on the real line $z=\oz$. 
More precisely, 
the general prescription of the boundary deformation
in  given $BCFT$ is as follows. We start with 
some $BCFT$ with the state space $\mathcal{H}^H_{\Omega,\alpha}$
where $(\Omega,\alpha)$ denotes the boundary 
condition along the real line. Boundary operators
$\psi(x)\in \Phi(\mathcal{H}^H)$ may be used to define a  
new perturbed $BCFT''$ whose correlation functions are constructed
from the unperturbed ones by the formal expansion
\begin{eqnarray}\label{def}
\left<\phi_1(z_1,\oz_1)\dots \phi_N(z_N,\oz_N)\right>_{\alpha,
\lambda \psi}=Z^{-1}\left<I_{\lambda \psi}
\phi_1(z_1,\oz_1)\dots \phi_N(z_N,\oz_N)\right>_{\alpha}=\nonumber 
\\ 
=Z^{-1}\sum_n\lambda^n \int \dots \int_{x_i<x_{i+1}}
\frac{dx_1}{2\pi}\dots 
\frac{dx_N}{2\pi}\left<\psi(x_1)\dots \psi(x_N)
\phi_1\dots \phi_N\right>_{\alpha} \ ,
\nonumber \\
\end{eqnarray}
where $\lambda$ is  a real parameter. From the
second line it is clear that the symbol
$I_{\lambda \psi}$ in the first line should be
understood as a path ordered exponential of
the perturbing operator
\begin{equation}
I_{\lambda \psi}=P\exp \left(
\lambda S_{\psi}\right)\equiv
P\exp \left(\lambda \int_{-\infty}^{\infty}
\frac{dx}{2\pi}\psi(x)\right)
\ .
\end{equation}
We must mention that given expression is rather formal
and suffers from UV divergences and should be
regularized. For more detailed discussion, see
again \cite{Recknagel:1998ih}.
And finally 
(\ref{def})  defines  deformations of
bulk correlators only. If there are extra boundary
fields present in the correlation function, these formulas
have to be modified so that these boundary fields
are included in the path ordering. However for
special class the boundary deformations these
formulas simplify considerably 
\cite{Recknagel:1998ih}. 
In (\ref{def}) the boundary operator $\psi$ has
conformal dimension $h$. For $h\neq 1$ 
the perturbation will automatically introduce
length scale and we have to follow  the
renormalization group flow to come back $BCFT$.
However as was stressed in \cite{Recknagel:1998ih} 
all these general perturbations  have 
common feature that the new $BCFT$ is associated
to the same bulk $CFT$. As a conclusion, the boundary
perturbations can only induce changes of the boundary 
conditions. For marginal deformations with $h=1$ this
implies that the boundary deformation induces 
the change of the original $BCFT$ with the gluing condition
$\Omega$ to the new $BCFT''$ with the new gluing
condition $\Omega''$, where the precise form of the 
 $\Omega''$ depends on the nature of $\psi$  \cite{Recknagel:1998ih}.

Let us consider the deformation of correlators
that contain boundary fields as well. It was shown
in  \cite{Recknagel:1998ih} that (\ref{def}) admits 
for the obvious generalization 
\begin{eqnarray}\label{corbound}
\left<\psi_1(u_1)\dots\psi_M(u_M)
\phi_1(z_1,\oz_1)\dots \phi_N(z_N,\oz_N)\right>_{\alpha,
\lambda \psi}=\nonumber \\
=Z^{-1}\sum_n\frac{\lambda^n}{n!} \int_{-\infty}
^{\infty} \dots \int_{-\infty}^{\infty}
\frac{dx_1}{2\pi}\dots 
\frac{dx_N}{2\pi}\left<\psi(x_1)\dots \psi(x_N)
\psi_1\dots \psi_M\phi_1\dots \phi_N\right>_{\alpha} \ ,
\nonumber \\
\end{eqnarray}
if and only if the boundary fields $\psi_1,\dots, \psi_M $ are
local with respect to the perturbing field $\psi$
\cite{Recknagel:1998ih}, where two boundary fields
$\psi_1(x_1) \ , \psi_2(x_2)$ are said to be local if
\begin{equation}
\psi_1(x_1)\psi_2(x_2)=
\psi_2(x_2)\psi_1(x_1) \ , x_1<x_2 \ .
\end{equation}
This equation is supposed to hold after insertion into
arbitrary correlation functions and the right hand side to
make sense it is required that there exists a
unique analytic continuation from $x_1<x_2$ to
$x_1>x_2$. We also say that a boundary field $\psi(x)$ 
is called self local or analytic if it is mutually
local with respect to itself. For example, the OPE
of a self-local boundary field $\psi$ with conformal
dimension $h_{\psi}=1$ is determined up to a constant
to be
\begin{equation}
\psi(x_1)\psi(x_2)=\frac{K}{(x_1-x_2)^2}+
\mathrm{reg} \ , h_{\psi}=1 \ .
\end{equation}
After appropriate renormalization 
\cite{Recknagel:1998ih} the correlation function
(\ref{corbound}) can be written as 
\begin{eqnarray}\label{corbound2}
\left<\psi_1(u_1)\dots\psi_M(u_M)
\phi_1(z_1,\oz_1)\dots \phi_N(z_N,\oz_N)\right>_{\alpha,
\lambda \psi}=\nonumber \\
=\sum_n\frac{\lambda^n}{n!} \int_{\gamma_1}
^{\infty} \dots \int_{\gamma_n}
\frac{dx_1}{2\pi}\dots 
\frac{dx_N}{2\pi}\left<\psi(x_1)\dots \psi(x_N)
\tilde{\psi}_1\dots 
\tilde{\psi}_M\phi_1\dots \phi_N\right>_{\alpha} \ ,
\nonumber \\
\end{eqnarray}
where $\gamma_p$ is the straight line parallel to the
real axis with $\mathrm{Im} \gamma_p=i\epsilon/p$ and
where the fields 
$\tilde{\phi}_i$ are given
\begin{equation}\label{redefR}
\tilde{\psi}_i(u_i)=\sum_{n=0}^{\infty}
\frac{\lambda^n}{2^n n!}
\oint_{C_1} \frac{dx_1}{2\pi }
\dots \oint_{C_n}\frac{dx_n}{2\pi}
\psi_i(_i)\psi(x_n)\dots \psi(x_1) \ .
\end{equation}
Note that  boundary fields $\mbw$
from the chiral algebra $\mw$ are  local with respect
to themselves  to all other boundary and bulk fields
in the theory so that (\ref{corbound2}) can 
be applied to correlators involving
arbitrary bulk and boundary fields. Then we 
see that (\ref{defm}) is special case of
(\ref{corbound2}) with no bulk operators inserted 
 and that (\ref{redef}) has the same
form as (\ref{redefR}). Consequently  we can claim that
redefinition of fluctuation modes (\ref{redef}) in
$S'$ (\ref{actionWF2}) maps this action to 
the SFT action (\ref{actionWF2N}) that
describes SFT action  defined on background $BCFT''$. 
\section{Conclusion}\label{fourth}
In this  note we have studied the
solution of open bosonic SFT based on the 
existence of marginal operators $\mbw$ from the
chiral algebra  $\mathcal{W}$ of $BCFT$, where
$BCFT$ is the classical background on which
 given SFT 
is defined.  We have mainly focused  on 
the relation between the fluctuation field around
the classical solution and the deformed $BCFT''$
that Arieses from the original $BCFT$ by inserting
marginal interaction on the real line. 
We  have seen that after
an expansion of the string field around the
classical solution and its insertion to 
the original action we obtain the SFT action $S'$
that after redefinition of
fluctuation fields  is written using
$BCFT''$ correlators that are deformations
of the correlators in $BCFT$ through 
introduction of perturbation $\mbw(x)$ from
the chiral algebra $\mw$ on the real line.
In other words, we have shown that two string
field theory actions $S_1\ , S_2$ defined using
two $BCFT's$ , $BCFT_1\ , BCFT_2$ where these two
$BCFT's$   are related
through marginal deformations from the chiral algebra
(It is important that $\lambda$ is not infinitesimal.), 
are in fact an expansion of SFT action around
different classical solutions. We mean that this 
result could be considered as an additional evidence
of the  background independence of open bosonic string
field theory  
\cite{Sen:1990hh,Sen:1990na,Sen:1992pw,Sen:1993mh,Sen:1994kx}
even in case of general deformation
parameter $\lambda$. We also hope that
this result could be helpful for recent
application of SFT, for example for the
study of the rolling tachyon solution. 
We hope to return to this problem in future. 
\\
{\bf Acknowledgment}
I would like to thank Ulf Danielsson and Ulf Lindstr\"om
for their support in my research.
This work is partly supported 
by EU contract HPRN-CT-2000-00122.
    
    \end{document}